# Electric field funnels for guiding charged nanoparticles.
# Simple models for exact solutions

*Peter V Pikhitsa*

e-mail: peterpikhitsa@gmail.com

**Abstract**

Static electric fields can be configured to guide charged nanoparticles along the electric field lines. Some field line configurations can focus the nanoparticles to prescribed places thus acting as electric funnels. Despite the importance of funnels for nanoprinting , there is a lack of formulation and understanding. We present two solvable models in two dimensions (2D) as far as it allows one to use the complex plane for exact calculation of the field lines. One model deals with two fixed dipoles such that the external electric field tries to squeeze through between them thus forming a funnel. The other one deals with a couple of parallel conducting plates with a hole where the external field comes through also creating a funnel. The results obtained may be useful for configuring the electric field for nanoprinting engineering.

**Introduction**

Positively charged nanoparticles can follow the electric field lines (streamlines) of a static electric field and therefore can be guided to prescribed places, which is important for such applications as nanoprinting [1]. The geometry of the streamlines can be tailored with the help of distributed charges fixed on a polymer as well as with biased conductive plates. In such a way one can split all streamlines into two domains, separated by a surface called a separatrix. Positively charged nanoparticles being injected into one domain remain in this domain and get funneled into prescribed places. The complementary (particle free) domain plays a role of the repelling cushion [2]. The separatrix has two sides: one from the domain with incoming particles and the other from the cushion side. Taking the particle side we call the separatrix the funnel. The concept of an electric funnel for focusing needs clear understanding in order to achieve the narrowest funnel which size is not too much sensitive to fluctuations in control parameters. The problem is how to predict the funnel shape.

Here we consider the problem in 2D which allows us to use the complex plane [3] in the first place. We take two simple models that yield exact solutions for the field in the whole plane and predict the funnel shape. One model deals with the funnel produced by two fixed electric dipoles placed in the external field. This model allows us to consider the process of funnel opening in detail. The second model [4] is the model with the three parallel plates; one plate at zero potential is at plus infinity in vertical direction and two other plates are at the same potential while the middle plate has a hole in it. The electric lines are released from the lower plate and go through the hole to infinity. As we will see, this case produces a funnel that never closes. To our best knowledge, neither of the models has been solved before though the first model was formulated in the Supplementary to [2].

**Opening a micro field electric funnel between the fixed dipoles by an external macro field**

Consider two 2D dipoles, both oriented upward along y-axis and placed along x-axis at $x_1 = -\omega$ and $x_2 = \omega$ so that the distance between them is $2\omega = W$ (Fig. 1).

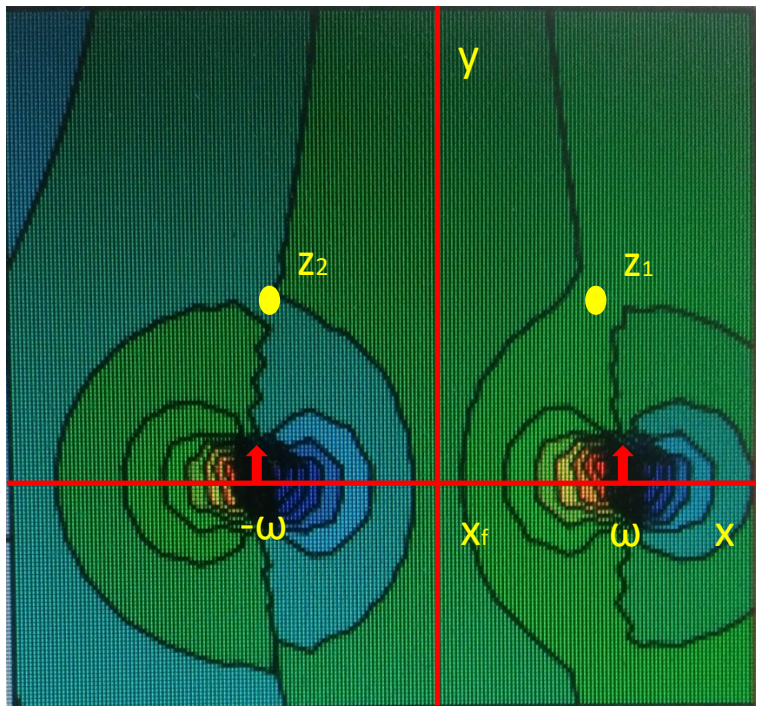

**Fig. 1.** Two dipoles producing an electric funnel of the half-width $x_f$ in a homogeneous electric field.

Let the macro-field $E$ be directed downward along y-axis. The separatrices between the field lines issuing from the dipoles and the lines of the macro-field $E$ define the edges of the electric funnel in the situation when the macro electric field lines can penetrate *between* the dipoles, thus opening and forming the funnel (see ref. [2] Supplementary). The stagnation equation ($-\frac{d\varphi}{dz} = 0$) defines two critical points $z_1$ and $z_2$ in the upper half of the plane (Fig. 1). The separatrix line passing through critical point $z_1$ follows the streamline function

Im $\varphi(z = z_1)$ = Im $\varphi(z)$ . (1)

The critical points of the complex potential for two dipoles (for simplicity we use here esu system and note that the dipole moment in 2D has the dimension of the charge [2])

$$\varphi(z) = \frac{-\tilde{q}i}{(z+\omega)} + \frac{-\tilde{q}i}{(z-\omega)} + zEi \quad (2)$$

are defined from the equation

$$-\frac{d\varphi}{dz} = \frac{\tilde{q}i}{(z+\omega)^2} + \frac{\tilde{q}i}{(z-\omega)^2} + Ei = 0 \tag{3}$$

which can be rewritten as a biquadratic equation

$$(z^2 - \omega^2)^2 + 2r^2(z^2 + \omega^2) = 0$$

with $r = \sqrt{\frac{\tilde{q}}{E}}$. The solution reads

$$z = \pm\sqrt{\omega^2 - r^2 \pm r\sqrt{r^2 - 4\omega^2}} \tag{4}$$

and we will consider only solutions in the upper right quarter of the plane Im $z > 0$, Re $z > 0$ because of the symmetry. In Fig. 2 we plot the evolution of the critical points with $E$. We see that at field $E < E_c = \frac{\tilde{q}}{4\omega^2}$ (that is $r = \frac{2\omega}{\sqrt{\frac{E}{E_c}}} > 2\omega$) we have two points lying along y-axis that do not "allow" the funnel to open. When the field $E \to 0$ ($r \to \infty$) the two critical points lie at $z_1 = ir\sqrt{2} \to i\infty$ and $z_2 = i\omega$. When the macro field increases the points approach one another and at $E = E_c$ they merge into one at $z_c = i\omega\sqrt{3}$ (see Fig. 2). For $E > E_c$ the critical points separate and leave y-axis. The funnel mouth is open now and the macro field lines penetrate the dipoles. At large fields $E \gg E_c$ ($r \ll \omega$) one can get from Eq. (4)

$$z = \pm\omega \pm ir \qquad (5)$$

so that the critical points approach individual dipole positions.

Let us write down the separatrix equation (1) with the critical point $z$ from Eq. (4) to find the half size of the funnel $x_f$ at $y = 0$

$$Im\ \varphi(z) = Im\left(\frac{-i4\omega^2 E_c 2z}{z^2 - \omega^2} + zEi\right) = Im\left(\frac{-i4\omega^2 E_c 2x_f}{x_f^2 - \omega^2} + x_f\ Ei\right). \qquad (6)$$

Unfortunately, it is a cubic equation with respect to $x_f$ so we do not have a simple answer at arbitrary field strength. However, the answer is simple for the fields in the vicinity of the critical field $E_c$ that is for the narrowest funnels at the brink of closure.

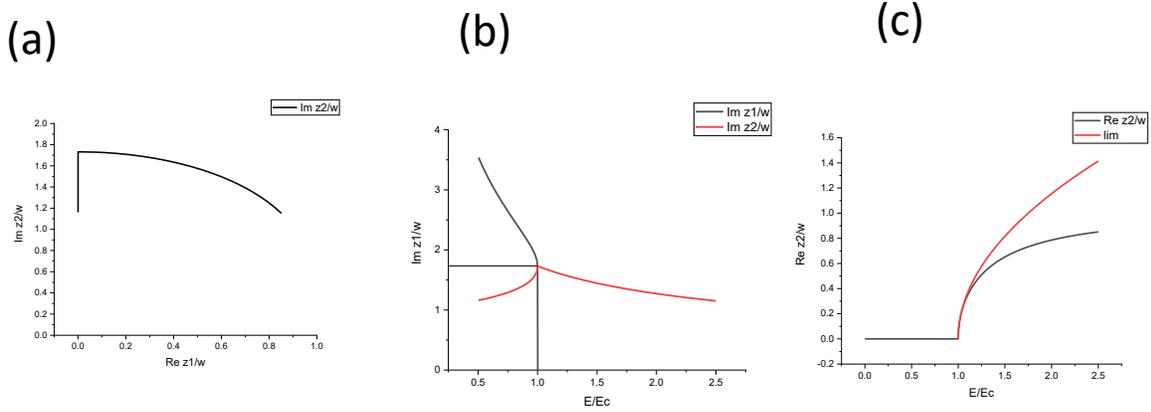

**Fig. 2:** Stagnation points on the external field. **a**, The evolution of the stagnation points over two dipoles depending on the external vertical field; **b**, Vertical position of the stagnation points on the external field strength. Red line corresponds to the right dipole for $E > E_c$ and to the lower stagnation point at $E < E_c$. **c**, Horizontal position of the stagnation points on the external field strength. The red line is from Eq. (11).

Let us study the funnel opening in the vicinity of the critical field $E_c$. Introduce

$$r = \frac{2\omega}{\sqrt{1+\delta}}, \qquad (7)$$

where $\delta = \frac{E}{E_c} - 1 \ll 1$. One can find from Eq. (4) that in this approximation

$$z_{1,2} \approx \sqrt{3}i\omega \mp \frac{2}{\sqrt{3}}\sqrt{\delta}\omega \,, \quad (8)$$

so that the square-root parabola of the real part of Eq. (8) matches the solution of Eq. (4) as is seen in Fig. 2c in the limit of small $\delta$.

Now we can analyze the separatrix equation (1) to find the half size of the funnel $x_f$ at $y = 0$ and $x_{f\infty}$ at $y = \infty$.

$$Im\, \varphi(z_2) = Im\left(\frac{-i4\omega^2 E_c}{z_2-\omega} + \frac{-i4\omega^2 E_c}{z_2+\omega} + z_2 E_c i\right) \approx Im\left(\frac{-i4\omega^2 E_c 2\frac{2}{\sqrt{3}}\sqrt{\delta}\omega}{-4\omega^2} + \frac{2}{\sqrt{3}}\sqrt{\delta}\omega E_c i\right) = 3\frac{2}{\sqrt{3}}\sqrt{\delta}E_c\omega \,. \quad (9)$$

On the other hand, at $z = x_f + 0i$ and $x_f \ll \omega$ we find

$$Im\, \varphi(z) \approx Im\left(\frac{-i4\omega^2 E_c * 2 * x_f}{-\omega^2} + x_f E_c i\right) = 9x_f E_c \,. \quad (10)$$

Equating Eq. (9) and (10) we obtain

$$x_f = \frac{2}{3\sqrt{3}}\sqrt{\delta}\omega = \frac{2}{3\sqrt{3}}\omega\sqrt{\frac{E}{E_c} - 1} \,. \quad (11)$$

The square-root dependence demonstrates a strong sensitivity of the focusing to the value of the applied field at extreme focusing in the vicinity of the critical field. Such a sensitivity may turn detrimental for a precise focusing and asks for an engineering of a stepwise focusing process. Finally, at $z = x_{f\infty} + \infty i$ we have

$$Im\, \varphi(z) = x_{f\infty} E_c \approx 9x_f E_c \,, \quad (12)$$

which demonstrates maximum 9 times squeezing of the global field flux into the funnel.

Note, that Eq. (11) can be even used for rough estimations for a flat conductive mask having a voltage difference $\Delta V$ with a flat conductive substrate, so that the surface charge density $\sigma = \Delta V/(4\pi d)$ and then (according to [2]) the dipole moment is $\tilde{q} = \sigma dr_d = \Delta V r_d/4\pi$, where $r_d$ is the Debye radius of the nanoparticle cloud.

### Electric field funnel in conformal mapping

Dipoles can be considered as an example of the fixed charge funnels. Yet, metal plates with movable charges may be used for focusing as well. Let us first consider the conformal mapping of the upper half complex plane to the horizontal strip with a cut along the negative abscissa (Fig.

3a). The complex potential for this mapping is derived in [3] (and illustrated in Fig. 3a, where the equipotential lines are shown):

$$\omega = \frac{1}{\pi}\left[h_1 \ln(1-z) + h_2 \ln\left(1 + \frac{h_1}{h_2}z\right)\right]. \quad (13)$$

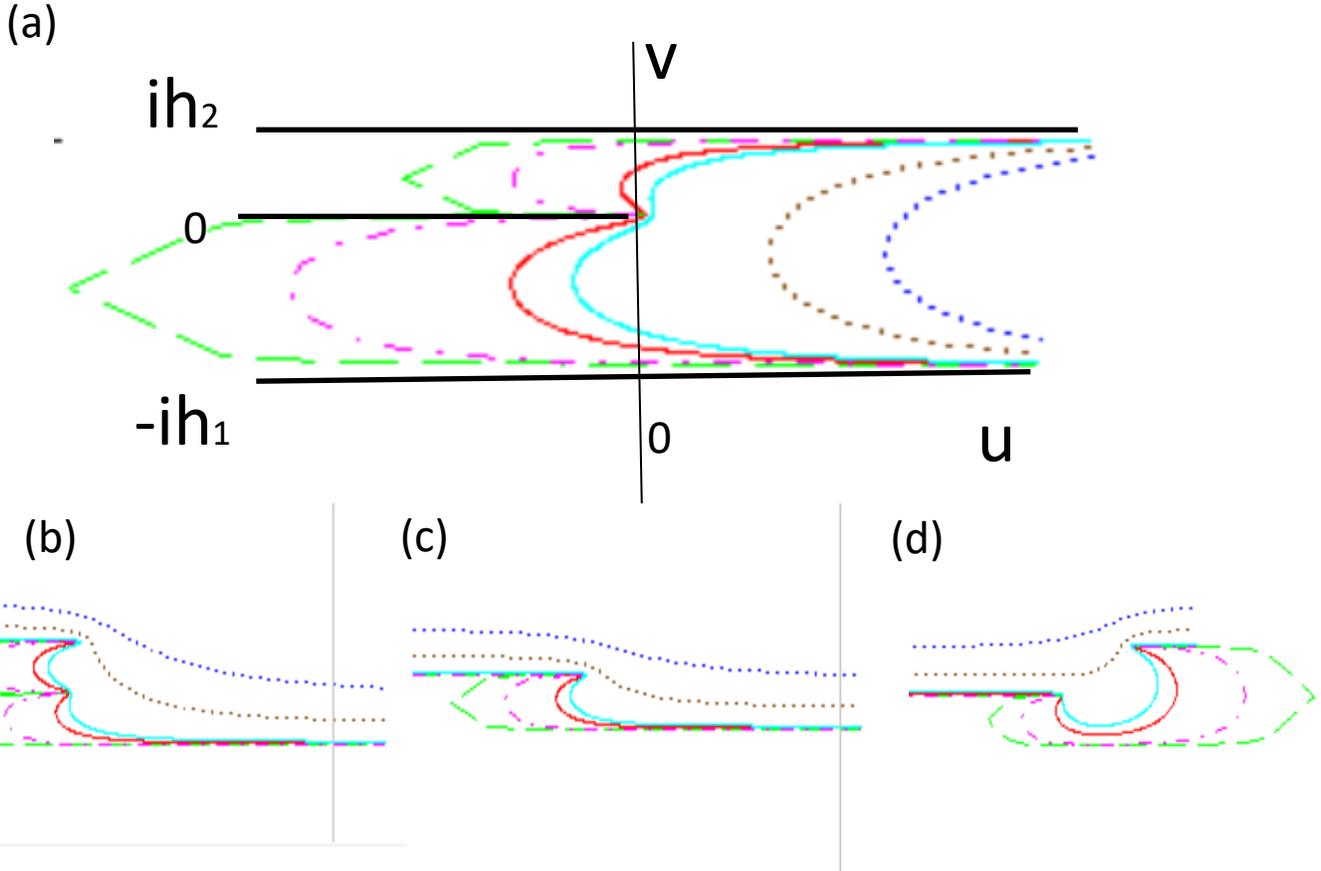

**Fig. 3**. Equipotential lines for various types of complex potentials (see text).

Manipulating parameters in Eq. (13) leads to three different cases depicted in Fig. 3b,c,d. First, let us set $h_2$ to infinity. Then Eq. (13) takes the form

$$\omega = \frac{1}{\pi}[h_1 \ln(1-z) + h_1 z]. \quad (14)$$

The potential lines are depicted in Fig. 3c. One can see that the linear term produces an external electric field which interacts with the plates. Therefore by just adding a linear function to Eq. (13) one obtains Fig. 3b. Then, by changing the sign in the second term one can get Fig. 3d that shows a hole. Now we are equipped with a general form of potential with two logarithmic and one linear terms to describe the simplest symmetric capacitor with a hole.

Let us consider 2D capacitor with a hole depicted in Fig.4b below.

(a)

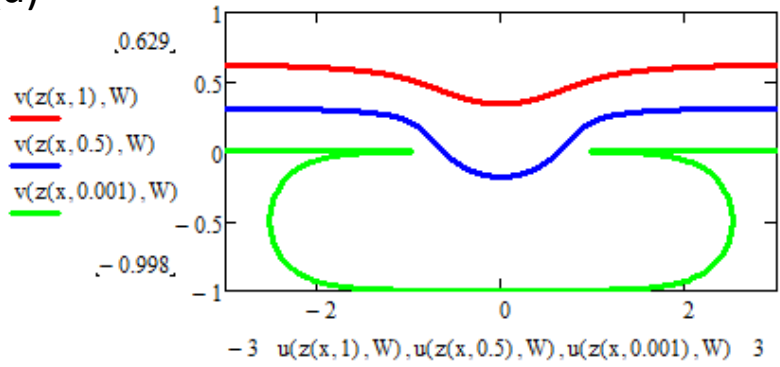

(b)

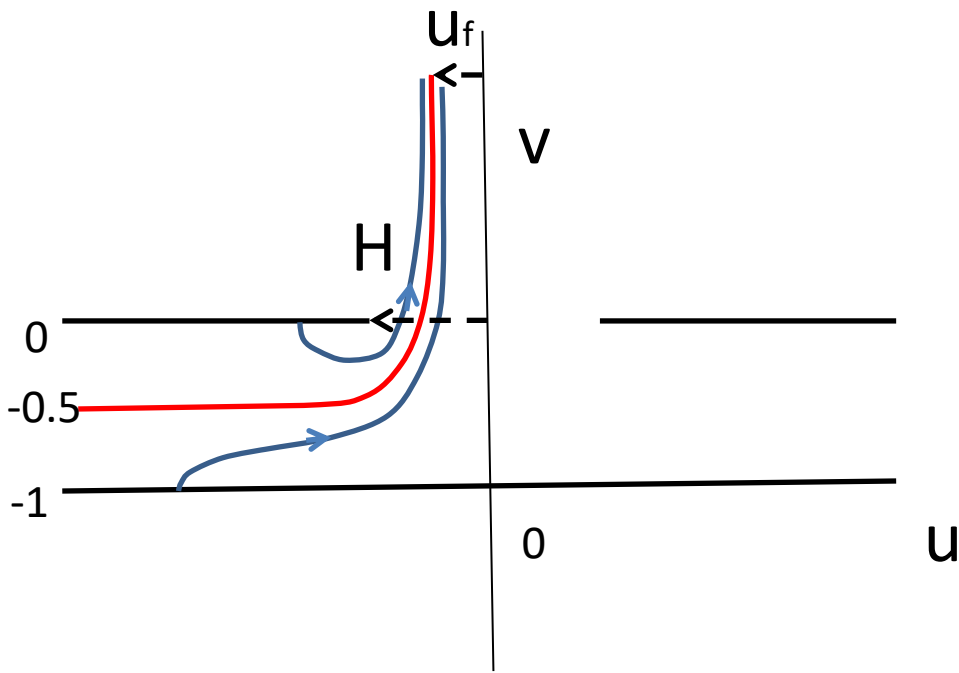

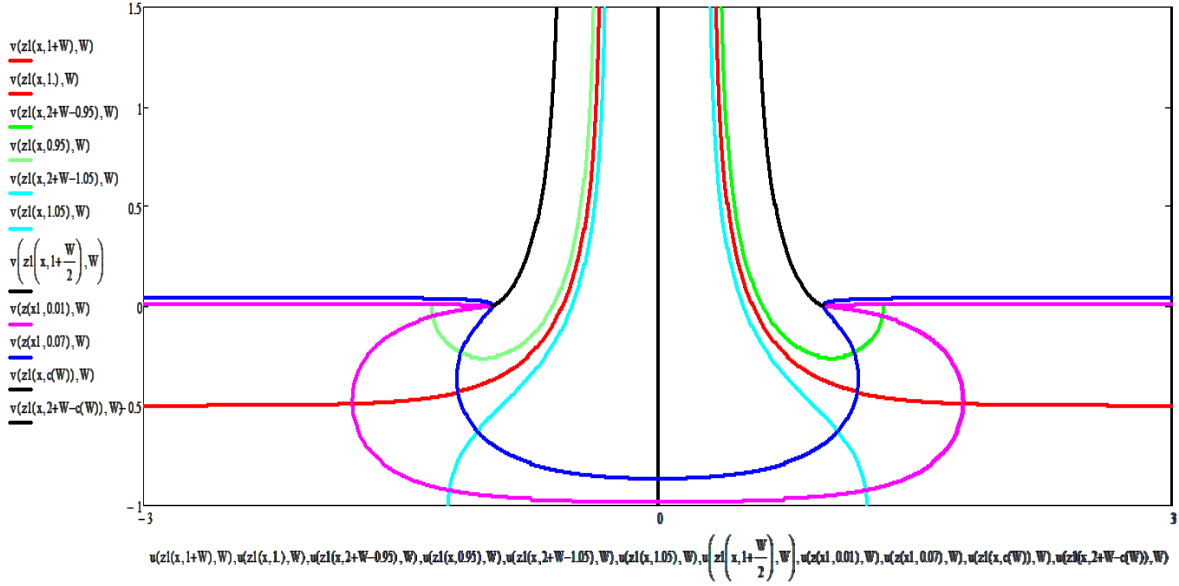

**Fig.4. a**, Equipotential lines for two equipotential plates with one hole in an external homogeneous electric field. **b**, Schematic of the field lines issuing from upper and lower plate divided by the separatrix (red line). **c**, Calculated field lines and equipotential lines at several parameters keeping vertical mirror symmetry. The funnel is the domain below the red lines. The secondary funnel is marked with black curved field lines.

The position of the hole in the upper plate is set symmetric with respect to the origin of the complex plane. The upper plate is at vertical coordinate 0 and the bottom plate is at -1 so that the distance between the plates is 1. Let both plates be positively charged and having the same potential. Following the method of conformal mapping described in [3], the three-term complex potential can be written as:

$$\omega = \frac{1}{\pi}[\ln(1-z) - \ln(1+W-z) + 2z - A]\, , (15)$$

where the shift $A$ is found from the condition that the equipotential lines are symmetric with respect to the origin. Defining complex

$$z = x + ic \quad (16)$$

we obtain the vertical coordinate $v$ and the horizontal coordinate $u$ of an equipotential line in a parametric form for $x$ running from minus to plus infinity, while keeping $c$ constant:

$$v(x) = Im(\omega) \quad (17)$$
$$u(x) = Re(\omega) \quad (18)$$

For the symmetric hole the line should be horizontal (that is $\frac{dv}{dx} = 0$) exactly at $u(x) = 0$. These two conditions give two equations so that the first one gives $x_0 = 1 + W/2$, while the second one (after substitution $u(x_0) = 0$) determines $A = 2 + W$. A set of equipotential lines is given in Fig. 4a for $W = 1$ and for $c = 1, 0.5$ and $0.001$.

The field lines are defined from the same parametric Eqs. (17) and (18), only now we must set $z = -ix + c$. Let us determine the separatrix field line and its asymptotic at infinity which gives the funnel size. When $c < 1$ a field line starts from the left part of the upper plate, while when $1 + W/2 > c > 1$ a field line starts from the left part of the bottom plate. The case $c = 1$ is the separatrix (see below). When $c = 1 + W/2$ the field line becomes vertical and for $c > 1 + W/2$ it starts on the right part of the bottom plate. The right part of the separatrix is at $c = 1 + W$. There is vertical mirror symmetry between the field lines determined by $c_{left}$ and $c_{right}$ when the sum $c_{left} + c_{right} = 2 + W$. To make a symmetric plot we chose the parameters for field lines in Fig. 4c accordingly to this rule. One can see that parameters for field lines sum up to 3 because we chose $W = 1$.

The set of field lines at various $c$ is depicted schematically in Fig. 4b for the left part of the funnel. The left part of the separatrix (left red line in Fig. 4c) is the electric field line at exactly $c = 1$, so that from (17) and (18) we have the parametric equation for the separatrix:

$v(x) = \frac{1}{\pi} Im[\ln(ix) - \ln(W + ix) - 2ix - W]$ (19)
$u(x) = \frac{1}{\pi} Re[\ln(ix) - \ln(W + ix) - 2ix - W]$ (20)

As one can see when $x$ approaches 0 from below then $u$ turns to minus infinity and $v$ approaches $-0.5$, which is exactly the half between the plates as it should be for the two equally charged plates. On the other side, when $x$ turns into minus infinity, the vertical coordinate $v$ goes to plus infinity while the horizontal coordinate $u$ approaches

$u_f = -W/\pi$ (21)

which determines the half of the funnel size. The funnel edge (separatrix) separates all the lines issuing from the upper plate with the hole, from the lines issuing from the bottom plate.

For completeness let us consider a "secondary" funnel (Fig. 4c), which side is a separatrix between electric field lines coming out of the hole and the lines issuing from the upper side of the upper plates. The left separatrix starts from the left edge of the hole with zero derivative, that is along the plate. From Eq. (17) rewritten with $z = -ix + c$ one has $v(x) = 0$ at $x = 0$. The derivative $\frac{dv}{du} = 0$ means that also $\frac{dv}{dx} = 0$ at $x = 0$. From Eq. (17) we obtain then a quadratic equation for :

$\frac{1}{1-c} - \frac{1}{1-c+w} = 2$,

which solution is

$$c = 1 + \frac{W}{2} \mp \frac{1}{2}\sqrt{2W + W^2},$$

so that the minus sign should be chosen for the left separatrix. Now with $x$ running to minus infinity $v$ runs to infinity and according to Eq. (18) with $z = -ix + c$ the horizontal coordinate approaches

$$u_{2f} = \frac{1}{\pi}(2c - 2 + W) = -\frac{1}{\pi}\sqrt{2W + W^2}$$

which determines the half size of the secondary funnel. One can compare this result with Eq. (21) to see that for narrow holes when $W \to 0$ the half of the secondary funnel becomes $\frac{-1}{\pi}\sqrt{2W} + \mathcal{O}(W)$, that is much broader than the first funnel and occupies the half of the hole size as we will see below from Eq. (28).

To find the size of the hole in the conformal mapping Eq. (15) we should note that the equipotential line at $c = 0$ fits the shapes of both plates (see Fig.4a for c=0.001). The hole edge can be found from the condition that the parametric equipotential line of Eqs. (17) and (18) has the infinite tangent at the edge, that is

$$\frac{dv}{du} = \frac{\frac{dv}{dx}}{\frac{du}{dx}} = \infty \,. \quad (22)$$

Eq. (22) gives the condition

$$\frac{du}{dx} = 0 \,. \quad (23)$$

An easy calculation with Eq. (18) and $c = 0$ gives

$$\frac{du}{dx} = \frac{1}{\pi}\left[\frac{-1}{1-x} + \frac{1}{1+W-x} + 2\right] = 0 \,. \quad (24)$$

After substituting one of the solutions of Eq. (24)

$$x_{1,2} = 1 + \frac{W}{2} \mp \frac{1}{2}\sqrt{2W + W^2} \quad (25)$$

into Eq. (18) we get the coordinate of the left edge of the hole:

$$u_h = \frac{-1}{\pi}\left[\ln\left(\frac{W+\sqrt{2W+W^2}}{-W+\sqrt{2W+W^2}}\right) + \sqrt{2W + W^2}\right] \,. \quad (26)$$

The ratio of the funnel size to the hole size

$$f = u_f/u_h \quad (27)$$

vs the absolute value $H = |u_h|$ is given in Fig.5. As far as the distance between plates was taken to be 1, then $H$ is equal to the ratio of the half of the hole size to the distance $d$ between plates for arbitrary distances. Finally, in the dimensional form the funnel size is $F = fh$ and the hole size $h = 2Hd$ while using the same parametric form for $f$ and $H$. Factor 2 appeared as far as the hole size is of two halves.

Now let us see what happens with the funnel size $F$ when the hole size $h$ shrinks to zero. This happens when $W \to 0$. Rewriting Eq.(26) we get

$$u_h = \frac{-1}{\pi}\left[\ln\left(1 + W + \sqrt{2W + W^2}\right) + \sqrt{2W + W^2}\right] \approx \frac{-1}{\pi} 2\sqrt{2W} + \mathcal{O}(W) \,. \quad (28)$$

That is $H \approx \frac{1}{\pi} 2\sqrt{2W}$ and $f = \frac{u_f}{u_h} \approx \frac{W}{2\sqrt{2W}} = \frac{\sqrt{W}}{2\sqrt{2}} = \frac{\pi}{8} H$. The universal proportionality factor of $\frac{\pi}{8}$ is shown in Fig. 5 with the black line. Thus, for the funnel size in the dimensional form we get

$$F = fh \approx \frac{\pi}{8} Hh = \frac{\pi}{16} \frac{h^2}{d} \,. \quad (29)$$

Eq. (29) shows that the funnel size shrinks quadratically fast with the narrowing of the hole size. This simple formula can be used for engineering a focusing device and indicates a possibility for strong control over focusing. Note that for the secondary funnel $f_2 = \frac{u_{2f}}{u_h} \approx \frac{\sqrt{2W}}{2\sqrt{2W}} = 1/2$ in the same limit $W \to 0$ as we claimed above.

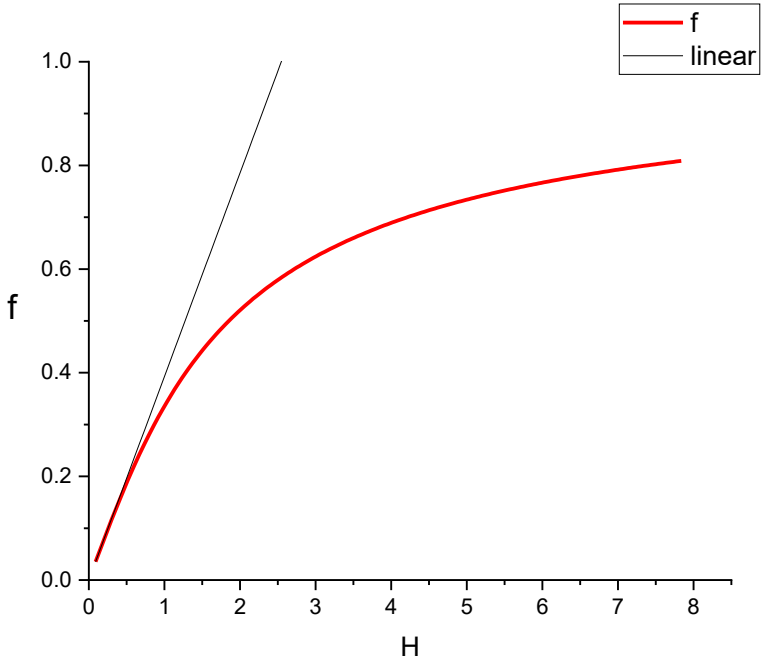

**Fig. 5**. Red line is calculated with Eqs. (26) and (27) in a parametric form. For small $H$ there is an exact linear approximation $f = \frac{\pi}{8} H$ (black line).

For practical applications it seems more convenient to combine Eq. (21) and (26) into one formula by substituting $W = \pi|u_f|$ from (21) into (26) and get a closed expression for the size of the hole $h = 2|u_h|d$ vs the size of the funnel $F = 2|u_f|d$ in the dimensional form:

$$h = \frac{2d}{\pi}\left[\ln\left(1 + \frac{\pi F}{2d} + \sqrt{\frac{\pi F}{d} + \left(\frac{\pi F}{2d}\right)^2}\right) + \sqrt{\frac{\pi F}{d} + \left(\frac{\pi F}{2d}\right)^2}\right].$$

For small funnels $\frac{\pi F}{2d} \ll 1$ this formula gives Eq. (29) as it should be.

Let us briefly discuss the magnitude of the electric field. It is clear that the complex potential of Eq. (15) defines the electric field distribution everywhere above the bottom plate. We present here only the analytical expression for the resulting magnitude of the normalized vertical electric field $E_v/E_0$ along the vertical symmetry line. On this line $z = 1 + \frac{W}{2} + ic$ (therefore $u \equiv 0$) and the vertical field is just a first derivative of the potential c over the vertical distance $v$ :

$$\frac{E_v}{E_0} = -\frac{dc}{dv} = -\left(\frac{dv}{dc}\right)^{-1}.$$

In parametric form:

$$\frac{E_v}{E_0} = \left[1 + \frac{2W}{4c^2 + W^2}\right]^{-1} , (30)$$

$$v = \frac{1}{\pi}\left[Im\left(\ln\left\{\frac{2ic+W}{2ic-W}\right\}\right) + 2c\right], (31)$$

where the running parameter $0 < c < \infty$ and $E_0$ is the magnitude of the flat field far away from plates. In Fig. 6 we plot the dependence. It is seen that on the bottom plate ($v = -1$) the field is $\frac{E_v}{E_0} = \frac{W}{W+2} < 1$ (that is the field is considerably reduced inside a narrow hole) and returns to 1 exponentially fast at the distance of approximately $2W$ above the upper plate. One can visually make sure from Fig. 4c that the ratio of the distance between the two equipotential lines above the upper plate away from the hole and the same two equipotential lines in the center close to the bottom plate is approximately 1/3 which corresponds to the value $\frac{W}{W+2}$ at $W = 1$. Note, that physically the increasing field on the centerline accelerates the charged nanoparticles through the hole while forming a jet. It happens because the viscosity leads to nanoparticle velocities proportional to local electric field.

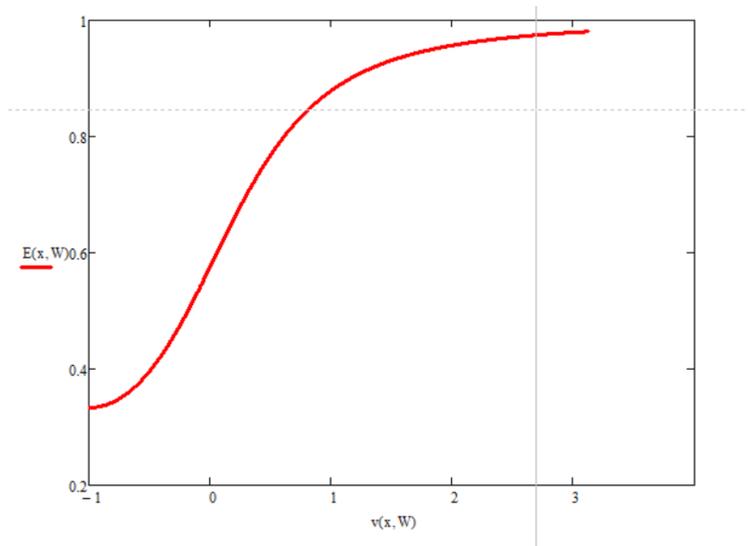

**Fig. 6**. Normalized electric field magnitude along the vertical centerline.

**Conclusion**

We presented exact solutions for two 2D models of electric funnels. One model with the fixed dipoles demonstrated that narrowing the funnel is not a simple task because of a square root dependence of the funnel size on the difference between the applied field and the critical field (see Eq. (11)). One can conclude that narrowing the funnel in one step is prone to focusing errors. The second model demonstrates that the funnel size can be quadratically (see Eq. (29)) narrowed with the manipulation with the sizes of conducting plates. The formulation can be easily extended for using more plates and their combinations for engineering. It also prompts the tactics of combining charges and/or plates with variable potentials for efficient and controllable focusing. However, an exact solution for such a general case is not yet found. Another interesting future problem is the shape of a 3D funnel.